# Green Supply Chain Management Optimization Based on Chemical Industrial Clusters

Jihu Lei

fjl@topsoe.com

Haldor Topsoe (Beijing) Co., Ltd., Beijing 100025, China

**Abstract.** During the period after the pandemic, the chemical sector, which is a crucial component of national progress, encounters further obstacles. The industry has been affected by the pandemic, leading to a need for faster transformation and upgrading. Additionally, there is an increased demand for green and environmental protection in the chemical sector due to global climate change and the concept of sustainable development. During the time after the pandemic, there will be a continued trend of industrial clustering and scale development in the chemical sector. It is essential to prioritize optimization via green supply chain management in order to facilitate industrial upgrading. In order to achieve this objective, this study administered a questionnaire survey to gather reliable data. The collected data was then analyzed using software tools like SPSS and AMOS. The analysis aimed to investigate the influence of various factors, including regulatory compliance, green procurement, green manufacturing, green logistics, green sales, competitors, internal environmental protection, and cost control, on the awareness and implementation of green supply chain management in chemical enterprises. Additionally, a structural equation model was constructed to further examine these relationships. The findings suggest that green procurement, manufacturing, logistics, sales, and internal environmental protection and cost control have a notable and beneficial influence on the implementation of green supply chain management. Additionally, the awareness and implementation of green supply chain management also contribute positively to both economic and environmental advantages. This paper offers a theoretical framework for enhancing the efficiency of the green supply chain management system in chemical industrial clusters, hence promoting the environmentally friendly and sustainable growth of the chemical sector.

**Keywords:** Green Supply Chain Management；Chemical Industrial Clusters；Green and sustainable development

## Introduction

In the post-pandemic era, the chemical industry - a pillar of national development - faces new challenges. On one hand, the industry's growth was impacted during the pandemic, necessitating an acceleration in transformation and upgrading. On the other hand, with the deepening of global climate change and the strengthening of the concept of sustainable development, there is a heightened demand for green and environmental protection measures in the chemical sector. Post-pandemic, the trend of industrial clustering and scale expansion in the chemical industry will continue, and optimization based on green supply chain management is crucial for promoting industrial upgrading. During the pandemic, supply chain disruptions led to severe challenges such as shortages of raw materials and logistics obstacles for the chemical industry, highlighting the need to build a green, sustainable, and stable supply chain. The path of large-scale development, park-style clustering, and centralized facilities is characteristic of chemical industrial clusters, which are formed by the integration of multiple supply chains. Due to increasingly stringent regulations, the chemical industry's exports face the issue of "green barriers," making the optimization research of green supply chain management based on chemical industrial clusters of significant importance. Drawing on experts' opinions, Chen., et al. [1] call for chemical companies to intensify the research, development, and application of green chemistry and clean production technologies, to reduce pollution emissions at the source [2]. Desing and Yu., et al. [3, 4] focus on the recycling of resources, suggesting that chemical companies strengthen cooperation with waste treatment enterprises to maximize resource reuse. Negri., et al. emphasize the need to consider from a holistic supply chain optimization perspective, enhance coordination among internal functional departments, and establish long-term partnerships with upstream and downstream enterprises to form a synergistic supply chain environmental management [5]. In the field of biochemistry, establishing an efficient and sustainable system is crucial [6, 7]. Recent advancements in biomedicine show that combining modern algorithms with efficient systems significantly boosts research output [8-11]. This not only accelerates the research process but also promotes the advancement of various sub-disciplines [12-14]. Big data simplifies analyzing large datasets, aiding in optimizing systems like traffic management for increased efficiency [15, 16]. Narwane., et al. [17] pay attention to the construction of supply chain informatization and intelligence, suggesting that chemical enterprises leverage new technologies such as big data and artificial intelligence to improve supply chain transparency and operational efficiency [18-20]. Currently, big data and artificial intelligence technologies have been widely applied across various research fields such as finance [21], structural health monitoring [20, 22], mechanical design [23, 24] and psychology [25, 26], achieving significant results. Particularly, their application in the biomedical field has been especially notable for its effectiveness [27-29]. In the post-pandemic era, although the impact of artificial intelligence and big data on students' learning [30-33], food imports [34] and sentiment analysis [35, 36] become a focal point of research, studies on the practical application and consequences in the supply chain field remain relatively scarce. Based on the analysis of the existing research, it appears that while previous studies have provided valuable insights and suggestions for optimizing green supply chain management in the chemical industry from various perspectives such as green chemistry, resource recycling, supply chain collaboration, and supply chain digitalization, there is still a lack of comprehensive empirical research specifically focusing on the influencing factors and optimization path of green supply chain management practices in the context of chemical industrial clusters in the post-pandemic era. Therefore, to address this research gap, we conducted a questionnaire survey to collect data from chemical enterprises located in representative chemical industrial clusters. By analyzing the collected data using structural equation modeling, this study empirically examined the impact of key factors such as regulatory compliance, green procurement, green manufacturing, green logistics, green sales, competitors, and internal environmental protection and cost control on the recognition and implementation of green supply chain management practices in chemical enterprises. Accordingly, a questionnaire survey was conducted to collect valid data, and data analysis software such as SPSS and AMOS were used to analyze the data to identify the main factors affecting the practice of green supply chain management in the chemical industry and to propose optimization suggestions.

# 1 Study on the Current Status and Influencing Factors of Green Supply Chain Management in the Chemical Industry

## 1.1 Questionnaire Design

In order to have a comprehensive understanding of the current situation and influencing factors of green supply chain management in chemical enterprises, a questionnaire was designed in this study. The survey questionnaire mainly consists of 10 sections: basic information of the company, regulatory compliance, green supply chain management practices, green procurement, green manufacturing, green logistics, green sales, competitors, internal environmental protection and cost control, as well as economic and environmental benefits [37]. Apart from the basic information that needs to be filled out, the measurement of all indicators in the remaining nine sections adopts a Likert five-point scale. In addition to basic information questions, a total of 49 questions were set up in the questionnaire survey. The focus of this study is to identify the main factors affecting the implementation of green supply chain management in chemical enterprises, including external regulatory pressures and competitors' behavior, as well as the environmental practices of internal departments such as purchasing, production, logistics, sales, etc., and the overall awareness of environmental protection and cost control in the enterprise. Except for the basic information, all the indicators in the nine sections were measured on a five-point Likert scale, with respondents scoring the extent to which the descriptions conformed to the actual situation of the enterprise, with 1 indicating very poor conformity and 5 indicating very good conformity. In addition to the basic information questions, the questionnaire set a total of 49 questions, covering all aspects of green supply chain management. Table 1 lists the specific measurement indicators of each dimension of the questionnaire. Regarding the regulatory compliance dimension, FG1 represents the extent to which a company complies with national environmental regulations, while FG2 and FG3 indicate the degree of compliance with local environmental regulations, which may involve different aspects of regulatory requirements. For the green procurement dimension, CG1 to CG4 represent the extent to which factors such as cooperating with suppliers to develop environmentally friendly technologies, selecting suppliers with environmental management system certification, and others are considered when choosing suppliers. In the green manufacturing dimension, ZZ1 and ZZ2 indicate the degree of cooperation between the company and upstream and downstream enterprises, ZZ3 to ZZ8 represent the emission and treatment of various wastes and exhaust gases during the production process, ZZ9 and ZZ10 indicate the company's ability to handle waste, while ZZ11 to ZZ15 reflect the company's performance in the construction of ecological industrial parks, including waste recycling, cooperation with upstream and downstream enterprises, and others. For the green logistics dimension, WL1 and WL2 represent the extent to which the company emphasizes environmental protection in logistics management, while WL3 and WL4 indicate the company's performance in resource reuse, improving transportation safety and efficiency, and so on. In the green sales dimension, XS1 to XS3 reflect the degree to which the company cooperates with customers to promote clean production, green packaging, energy-saving transportation, and others during the sales process, while XS4 to XS7 represent the company's recycling and reuse of various resources and waste. These specific indicators are collected through questionnaire items to measure the company's practices and performance in various aspects of green supply chain management, thereby scoring each influencing factor. Through the measurement of these indicators, the awareness and practice level of chemical enterprises in green supply chain management can be comprehensively assessed, and the weaknesses can be identified to lay the foundation for the subsequent optimization suggestions.

**Table 1 Influencing Factors of Green Supply Chain Management in the Chemical Industry**

| Dimension | Factor | Specific measurement indicators |
|---|---|---|
| Regulatory Compliance | National Environmental Regulations | FG1 |
| | Local Environmental Regulations | FG2-FG3 |
| Green Procurement | Supplier Selection | CG1-CG4 |
| | Strengthening Relationships with Suppliers to Co-develop New Technologies | |
| | Selecting Suppliers with ISO14001 Certification | |
| | Cooperation with Suppliers on Environmental Issues | |
| Green Manufacturing | Waste Recycling | ZZ1-ZZ13 |
| | Emphasis on Recycling Waste from Upstream and Downstream Enterprises as Raw Materials | |
| | Relationship with Upstream and Downstream Enterprises | |
| | Waste in Manufacturing | |
| | Enterprise Waste Disposal | |
| | Joint Technological Innovation with Upstream and Downstream Enterprises | |
| | Joint Industrial Park Development to Reduce Costs | |
| | Emissions from Combustion of Furnaces and Boilers | |
| | Non-condensable Gases from Production Units | |
| | By-products from Venting Gases and Reactions | |
| | Volatile Emission of Light Oil and Chemicals | |
| | Toxic Gases during Wastewater and Waste Treatment and Transportation | |
| | Exhaust Gases during Transportation of Raw Materials and Products in Chemical Production | |
| | Improvements in Chemical Production Processes and Equipment | |
| | A Well-established System to Strengthen Emission Management of Pollutants | |
| | Ecological Industrial Park to Reduce Transportation Costs | |
| | Ecological Industrial Park | |
| | Waste Recycling into Raw Materials for Other Products | |
| | Close Links between Upstream and Downstream, Joint Investment in New Technologies | |
| | Logistics Management | |
| | Resource Reuse | |
| Green Logistics | Emphasis on Safe Transportation to Avoid Corrosion and Leakage | WL1-WL4 |
| | Improving Transportation Equipment Quality to Reduce Hazards and Energy Consumption | |
| Green Sales | Sales Process | XS1-XS9 |
|  | Cooperation with Customers on Clean Production |  |
|  | Cooperation with Customers on Green Packaging |  |
|  | Cooperation with Customers on Reducing Product Transportation Energy Consumption |  |
|  | Recycling Process |  |
|  | Recycling and Selling Excess Inventory or Materials |  |
|  | Recycling and Selling Scrap or Used Materials |  |
|  | Recycling and Selling Idle Equipment Assets |  |
|  | Market Factors |  |
|  | Selling Products to Foreign/ Joint Ventures in China |  |
|  | Environmental Awareness among Chinese Consumers |  |
|  | Establishing a Green Corporate Image |  |
|  | Competitors' Green Strategies |  |
| Competitors |  | JD1-JD2 |
|  | Activities of Industry Associations |  |
|  | Environmental Factors |  |
|  | Parent Company's Environmental Strategy |  |
| Internal Environmental Protection and Cost Control |  | NY1-NY6 |
|  | Company's Environmental Vision (Goals) |  |
|  | Cost Factors |  |
| | Potential Liability for Disposal of Hazardous Materials | |
| | Disposal Costs of Hazardous Materials | |
| | Costs of Environmentally Friendly Products | |
| | Costs of Environmentally Friendly Equipment (Transportation, Packaging) | |
| | Recognition of Green Supply Chain Management | |
| | Implementing Green Supply Chain Management is Beneficial to Environmental Protection | |
| Green Supply Chain Management Practices | Implementation of Green Supply Chain Management | XZ1-XZ5 |
| | Building a Green Supply Chain Management System | |
| | Enterprise has Obtained ISO14001 Certification | |
| | Creating Positions and Departments Related to Green Supply Chain Management | |
| | Economic Benefits | |
| | Environmental Benefits | |
| Economic and Environmental Benefits | Reducing Costs and Increasing Benefits | XY1-XY3 |
| | Reducing the Degree of Environmental Pollution | |
| | Environmental Improvement, Reducing Environmental Management Costs | |

## 1.2 Sample Selection and Data Collection

This paper conducts its survey in two cities, Shanghai and Jinan, because Shanghai is home to relatively mature chemical industrial parks, while Jinan features comparatively younger chemical industrial parks. Additionally, the industrial parks in both Shanghai and Jinan share characteristics of significant size and a complex array of chemical enterprises. The survey questionnaires were distributed both on-site and online. A total of 150 questionnaires were disseminated, with 148 collected in return. Out of these, 147 were deemed valid. The analysis that follows is based on the data from these 147 valid responses.

### 1.2.1 Reliability and Validity Analysis

（1）Reliability Analysis

Reliability analysis refers to the assessment of the consistency of survey results, testing for stability and reliability of the data obtained. This paper utilizes SPSS 20.0 software to conduct a reliability analysis of the questionnaire, using Cronbach's alpha coefficient as the standard measure of reliability. The Cronbach's alpha coefficients for each factor, as well as for the overall scale, are presented in Table 2.

Table 2 Reliability Analysis using Cronbach's alpha coefficient

| Indicator | Factor | Factor Reliability | Indicator Reliability | Total Reliability |
|---|---|---|---|---|
| Regulatory Compliance | Regulatory Compliance | 0.898 | 0.898 | |
| Green Supply Chain Management Practices | Recognition of Green Supply Chain Management | 0.927 | 0.898 | |
| | Implementation of Green Supply Chain Management | 0.810 | | |
| Green Procurement | Supplier Selection | 1.000 | 0.772 | 0.775 |
| | Waste Recycling | 0.919 | | |
| Green Manufacturing | Upstream and Downstream Relationships | 0.945 | 0.850 | |
| | Waste in Manufacturing | 0.846 | | |
| | Enterprise Waste Disposal | 0.939 | | |
| | Ecological Industrial Park | 0.855 | | |

| | | | |
|---|---|---|---|
| Green Logistics | Logistics Management | 0.819 | 0.813 |
| | Resource Reuse | 0.906 | |
| Green Sales | Sales Process | 0.787 | 0.847 |
| | Recycling Process | 0.812 | |
| | Market Factors | 0.979 | |
| Competitors | Competitors | 0.954 | 0.954 |
| Internal Environmental Cost Control | Environmental Factors | 0.827 | 0.785 |
| | Cost Factors | 0.903 | |
| Economic and Environmental Benefits | Economic Benefits | 1.000 | 0.790 |
| | Environmental Benefits | 0.785 | |

From Table 2, it can be seen that the Cronbach's alpha for each factor is greater than 0.7, and the Cronbach's alpha for each indicator is also greater than 0.7. The total scale's Cronbach's alpha is 0.775, which is close to 0.8, indicating that the survey questionnaire has good stability.

1.2.2 Validity Analysis

This study employed SPSS 20.0 software to conduct factor analysis on the collected data to examine the construct validity of the questionnaire. The objects of analysis included nine constructs: regulatory compliance, green supply chain management practices, green procurement, green manufacturing, green logistics, green sales, competitors, internal environmental protection and cost control, as well as economic and environmental benefits.

Before performing factor analysis, the Kaiser-Meyer-Olkin (KMO) test and Bartlett's test of sphericity were utilized to assess the suitability of the data for factor analysis. Generally, a KMO value greater than 0.8 indicates strong correlations among variables, making them highly suitable for factor analysis; while a KMO value between 0.7 and 0.8 suggests relatively strong correlations among variables, indicating suitability for factor analysis. Bartlett's test of sphericity has a null hypothesis that the correlation matrix is an identity matrix, implying no significant correlations among variables. Therefore, if the significance level is less than 0.05, the null hypothesis should be rejected, indicating correlations among variables and suitability for factor analysis.

The analysis results showed that the KMO values for the nine constructs were 0.767, 0.785, 0.916, 0.833, 0.876, 0.889, 0.754, 0.813, and 0.757, respectively, all close to or exceeding 0.8. Meanwhile, the

probabilities derived from Bartlett's test of sphericity were all 0.000, which is significantly less than the 0.05 significance level. Consequently, the null hypothesis should be rejected, suggesting that the collected data is suitable for factor analysis.

After confirming the suitability of the data, factor extraction was performed, and several common factors were identified after factor rotation. Tables 3 to 11 respectively present the rotated factor loading matrices for regulatory compliance, green supply chain management practices, green procurement, green manufacturing, green logistics, green sales, competitors, internal environmental protection and cost control, and economic and environmental benefits, as well as the cumulative variance contribution rates of each factor.

**Table 3 Factor Loading Matrix for Regulatory Compliance Measurement**

| Indicator | Factor 1 |
| --- | --- |
| FG1 | 0.875 |
| FG2 | 0.883 |
| FG3 | 0.932 |
| Cumulative Variance Contribution Rate | 78.671% |

**Table 4 Green Supply Chain Management Factor Loading Matrix**

| Indicator | Factor 2 | Factor 3 |
| --- | --- | --- |
| XZ1 | 0.755 | 0.255 |
| XZ2 | 0.901 | 0.151 |
| XZ3 | 0.178 | 0.869 |
| XZ4 | 0.463 | 0.647 |
| XZ5 | 0.230 | 0.778 |
| Cumulative Variance Contribution Rate | 35.006% | 66.787% |

**Table 5 Green Procurement Factor Loading Matrix**

| Indicator | Factor 4 | Factor 5 |
|---|---|---|
| CG1 | 0.838 | -0.073 |
| CG2 | 0.812 | -0.072 |
| CG3 | 0.851 | 0.130 |
| CG4 | 0.030 | -0.795 |
| Cumulative Variance Contribution Rate | 44.843% | 86.601% |

**Table 6 Green Manufacturing Factor Loading Matrix**

| Indicator | Factor 6 | Factor 7 | Factor 8 | Factor 9 |
|---|---|---|---|---|
| ZZ1 | 0.933 | -0.202 | -0.031 | 0.030 |
| ZZ2 | 0.919 | -0.239 | -0.094 | 0.050 |
| ZZ3 | -0.063 | 0.871 | 0.043 | 0.118 |
| ZZ4 | -0.031 | 0.790 | 0.184 | 0.100 |
| ZZ5 | -0.048 | 0.862 | 0.035 | 0.144 |
| ZZ6 | -0.226 | 0.763 | -0.091 | 0.061 |
| ZZ7 | -0.158 | 0.821 | 0.069 | 0.150 |
| ZZ8 | -0.109 | 0.746 | 0.208 | 0.124 |
| ZZ9 | 0.003 | 0.143 | 0.751 | 0.555 |
| ZZ10 | -0.018 | 0.226 | 0.936 | 0.126 |

| Indicator | Factor 6 | Factor 7 | Factor 8 | Factor 9 |
| --- | --- | --- | --- | --- |
| ZZ11 | -0.148 | 0.187 | 0.055 | 0.876 |
| ZZ12 | 0.042 | 0.016 | 0.060 | 0.899 |
| ZZ13 | -0.004 | 0.128 | 0.123 | 0.891 |
| Cumulative Variance Contribution Rate | 12.247% | 52.023% | 61.132% | 82.906% |

**Table 7 Green Logistics Factor Loading Matrix**

| Indicator | Factor 10 | Factor 11 |
| --- | --- | --- |
| WL1 | 0.779 | 0.016 |
| WL2 | 0.924 | -0.130 |
| WL3 | -0.024 | 0.913 |
| WL4 | -0.090 | 0.922 |
| Cumulative Variance Contribution Rate | 48.856% | 80.956% |

**Table 8 Green Sales Factor Loading Matrix**

| Indicator | Factor 12 | Factor 13 | Factor 14 |
| --- | --- | --- | --- |
| XS1 | 0.808 | 0.079 | -0.284 |
| XS2 | 0.882 | 0.122 | -0.062 |
| XS3 | 0.898 | 0.092 | -0.169 |
| XS4 | 0.009 | 0.648 | 0.598 |
| XS5 | -0.075 | 0.735 | 0.512 |

| Indicator | Factor 12 | Factor 13 | Factor 14 |
|---|---|---|---|
| XS6 | 0.021 | 0.704 | 0.475 |
| XS7 | 0.306 | -0.331 | 0.858 |
| XS8 | 0.315 | -0.314 | 0.842 |
| XS9 | 0.314 | -0.332 | 0.857 |
| Cumulative Variance Contribution Rate | 41.884% | 55.594% | 82.468% |

**Table 9 Competitors Factor Loading Matrix**

| Indicator | Factor 15 |
|---|---|
| JD1 | 0.977 |
| JD2 | 0.979 |
| Cumulative Variance Contribution Rate | 95.633% |

**Table 10 Internal Environmental Protection and Cost Control Factor Loading Matrix**

| Indicator | Factor 16 | Factor 17 |
|---|---|---|
| NY1 | 0.791 | -0.307 |
| NY2 | 0.728 | -0.433 |
| NY3 | 0.801 | -0.403 |
| NY4 | 0.349 | 0.858 |
| NY5 | 0.351 | 0.858 |
| NY6 | 0.379 | 0.807 |

| Indicator | Factor 16 | Factor 17 |
|---|---|---|
| Cumulative Variance Contribution Rate | 36.736% | 79.174% |

Table 11 Economic and Environmental Benefits Factor Loading Matrix

| Indicator | Factor 18 | Factor 19 |
|---|---|---|
| XY1 | 0.955 | 0.242 |
| XY2 | 0.173 | 0.945 |
| XY3 | 0.507 | 0.728 |
| Cumulative Variance Contribution Rate | 40.209% | 89.605% |

From the aforementioned analysis, a total of 19 common factors have been extracted to represent the items in the survey questionnaire. Each factor is named according to the content of the items it encompasses. Factor 1 is named Regulatory Compliance, with a cumulative variance contribution rate of 78.671%. Factor 2 is named Recognition of Green Supply Chain Management, and Factor 3 is named Implementation of Green Supply Chain Management, with a cumulative variance contribution rate of 66.787%. Factors 4 and 5 are named Supplier Selection and Waste Recycling, respectively, with these two factors reaching a cumulative variance contribution rate of 86.601%. Factor 6 is named Upstream and Downstream Relationships, Factor 7 is named Waste in Manufacturing, Factor 8 is named Enterprise Waste Disposal, and Factor 9 is named Ecological Industrial Park, with these four factors having a cumulative variance contribution rate of 82.906%. Factors 10 and 11 are named Logistics Management and Resource Reuse, respectively, with a combined cumulative variance contribution rate of 80.956%. Factors 12, 13, and 14, named Sales Process, Recycling Process, and Market Factors, have a cumulative variance contribution rate of 82.468%. Factor 15, named Competitors, has a cumulative variance contribution rate of 95.633%. Factors 16 and 17 are named Environmental Factors and Cost Factors, with a cumulative variance contribution rate of 79.174%. Factors 18 and 19 are named Economic Benefits and Environmental Benefits, with a cumulative variance contribution rate of 89.605%. This indicates that the validity of the survey questionnaire is quite good.

## 2 Model Construction

### 2.1 Research Hypotheses

Based on the survey and analysis of the aforementioned information, the following hypotheses are proposed:

The cumulative variance contribution rate of the factor Regulatory Compliance is 78.671%, indicating that this factor can well explain the variation in the questionnaire responses regarding the corporate compliance with environmental regulations. This suggests that regulatory compliance has a significant impact on the implementation of green supply chain management by enterprises, leading to the proposition of Hypothesis H1.

H1: Regulatory compliance has a significant positive impact on the practice of green supply chain management in enterprises.

The cumulative variance contribution rate for the two factors of Supplier Selection and Waste Recycling in green procurement reaches 86.601%, indicating that these factors adequately reflect the green supply chain management practices in the procurement phase of enterprises. Based on this, Hypothesis H2 is proposed, positing that the implementation of green procurement has a significant positive impact on the green supply chain management practices of enterprises.

H2: The implementation of green procurement has a significant positive impact on the practice of green supply chain management in enterprises.

The four factors of green manufacturing (Supply Chain Relationships, Waste in Production Process, Corporate Waste Treatment, and Eco-Industrial Parks) have a cumulative variance contribution rate of 82.906%, indicating that these factors comprehensively reflect the green supply chain management practices during the production and manufacturing stages of enterprises. Consequently, Hypothesis H3 is proposed.

H3: The implementation of green manufacturing has a significant positive impact on the practice of green supply chain management in enterprises.

The two factors of green logistics (Logistics Management and Resource Reutilization) have a cumulative variance contribution rate of 80.956%, suggesting that these factors effectively reflect the green supply chain management practices within the logistics segment of enterprises. Based on this finding, Hypothesis H4 is proposed.

H4: The implementation of green logistics has a significant positive impact on the practice of green supply chain management in enterprises.

The three factors of green sales (Sales Process, Recycling Process, and Market Factors) have a cumulative variance contribution rate of 82.468%, indicating that these factors provide a comprehensive reflection of the green supply chain management practices in the sales phase of enterprises. Consequently, Hypothesis H5 is proposed.

H5: The implementation of green sales has a significant positive impact on the practice of green supply chain management in enterprises.

The cumulative variance contribution rate for the Competitor factor is as high as 95.633%, suggesting that this factor has a strong explanatory power for the variation in questionnaire responses regarding competitors' green strategies and industry association activities. Consequently, Hypothesis H6 is proposed, positing that competitors may have a negative impact on the green supply chain management practices of enterprises.

H6: Competitors have a significant negative impact on the practice of green supply chain management in enterprises.

The cumulative variance contribution rate for the two factors of Internal Environmental Protection and Cost Control (Environmental Factors and Cost Factors) is 79.174%, indicating that these factors effectively reflect the considerations of enterprises internally regarding environmental protection and cost. Based on this, Hypothesis H7 is proposed.

H7: Internal environmental protection and cost control have a significant positive impact on the practice of green supply chain management in enterprises.

The cumulative variance contribution rate for the two factors of Green Supply Chain Management Awareness and Implementation is 66.787%, while the cumulative variance contribution rate for the factors of Economic and Environmental Benefits reaches 89.605%. This indicates a close correlation between enterprises' recognition and execution of green supply chain management and their economic and environmental performance. Consequently, Hypotheses H8 and H9 are proposed, positing that awareness and implementation of green supply chain management have a positive impact on the economic and environmental benefits of enterprises.

H8: The recognition of green supply chain management has a significant positive impact on the economic and environmental benefits of enterprises.

H9: The implementation of green supply chain management has a significant positive impact on the economic and environmental benefits of enterprises.

**2.2 Structural modeling**

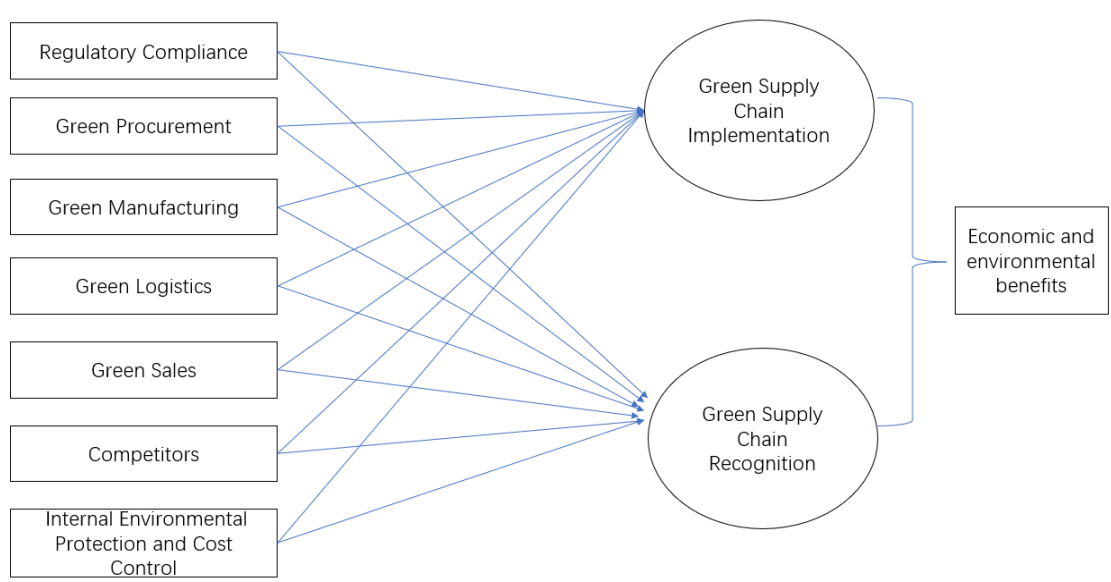

Figure 1: The Initial Structure Model

As shown in Figure 1, based on the relevant research hypotheses mentioned above, the initial structural model of the factors influencing the advancement of green supply chain management practices can be derived.

**2.3 Model Fitting and Modification**

Based on the initial structural model described above, AMOS 20.0 software was used to fit and modify the model to determine its accuracy. Table 12 provides the standardized path coefficients for the initial structural equation model.

**Table 12 Standardized Path Estimates for the Initial Model**

| Path | C.R. | P |
|---|---|---|
| Recognition of GSCM <- Regulatory Compliance | 2.478 | *** |
| Recognition of GSCM <- Green Procurement | -2.791 | *** |
| Recognition of GSCM <- Green Manufacturing | -3.231 | 0.001 |
| Recognition of GSCM <- Green Logistics | 2.348 | 0.019 |
| Recognition of GSCM <- Green Sales | 3.802 | *** |
| Recognition of GSCM <- Competitors | -1.821 | 0.805 |
| Recognition of GSCM <- Env. & Cost Control | -3.247 | *** |
| Implementation of GSCM <- Regulatory Compliance | 14.675 | *** |
| Implementation of GSCM <- Green Procurement | -5.781 | *** |
| Implementation of GSCM <- Green Manufacturing | -3.007 | 0.038 |
| Implementation of GSCM <- Green Logistics | 7.394 | 0.049 |
| Implementation of GSCM <- Green Sales | 2.876 | *** |
| Implementation of GSCM <- Competitors | -0.971 | 0.091 |
| Implementation of GSCM <- Env. & Cost Control | -8.997 | *** |
| Economic & Env. Benefits <- Recognition | -6.554 | *** |
| Economic & Env. Benefits <- Implementation | 6.110 | *** |

Note: GSCM stands for Green Supply Chain Management. In the 'P' column, "***" indicates a significance level of 0.001 or lower, and numerical values represent the p-value for the significance of the path coefficient.

In Table 12, the absolute C.R. values for the paths "Recognition of Green Supply Chain Management <-- Competitors" and "Implementation of Green Supply Chain Management <-- Competitors" are both less than 1.96, and the corresponding P-values are greater than 0.05. This does not satisfy the standard proposed by relevant experts, which states that the t-values of model path coefficients should be greater than 1.96. It indicates that the influence of competitors on enterprises' recognition and implementation of green supply chain management is not significant, and these influence paths need to be removed in the revised model. Eliminating insignificant influence paths helps to improve the model's goodness of fit and explanatory power. By gradually modifying or removing paths with t-values that do not meet the requirements, a more concise and explanatory model can be obtained. This helps to highlight the role of key influencing factors in green supply chain management and provides clearer and more actionable directions for chemical enterprises to optimize their green supply chain management practices. After the model revision, the t-values of each influence path satisfy the standard of being greater than 1.96, as shown in Figure 2. The revised model better reveals the significant positive influence of factors such as regulatory compliance, green procurement, green manufacturing, green logistics, green sales, as well as internal environmental protection and cost control on chemical enterprises' recognition and implementation of green supply chain management. It validates the hypotheses other than the one related to competitors. Meanwhile, the positive impact of green supply chain management recognition and implementation on enterprises' economic and environmental benefits is also supported by the data [38]. The revised model is presented in Figure 2.

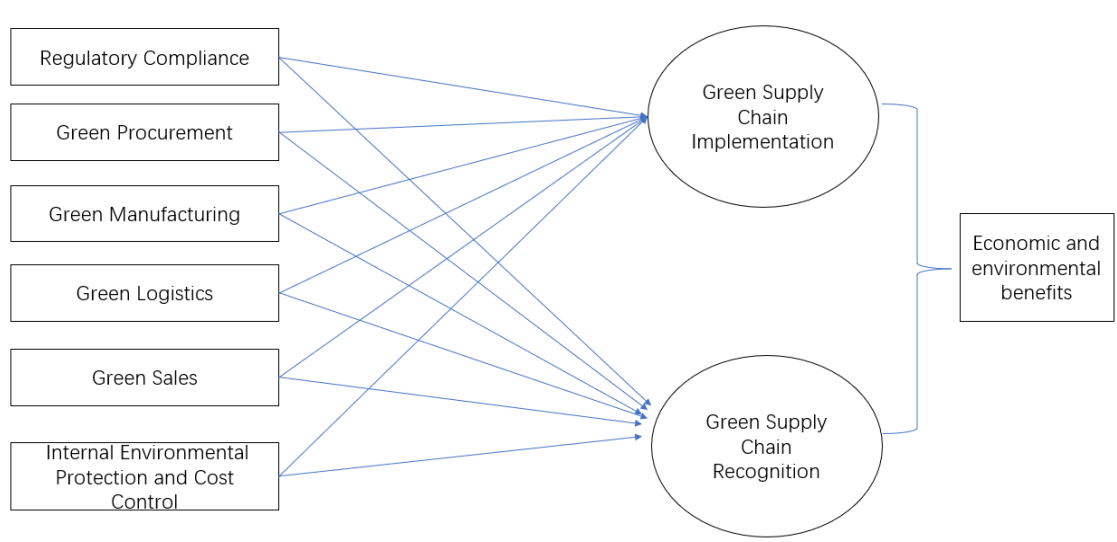

**Figure 2 modified model**

The revised model was tested using AMOS 20.0 software. Since there were 165 valid questionnaires from this survey, data from 165 respondents were obtained to determine whether the model, with a sample size of 165, required further modification. The primary focus was to check whether four fit indices had met the adequacy standards. The results showed that all indices met the standards, indicating that the model is relatively ideal. Table 13 presents the fit indices for the revised model.

**Table 13 Model Fit Indices**

| Index Standard | Test | Model Assessment |
|---|---|---|
| CMIN/DF | 1-3 | 1.737 Acceptable |
| RMSEA < 0.08 (Reasonable Fit) < 0.05 (Good Fit) | 0.049 | Acceptable |
| AGFI > 0.9 | 0.943 | Acceptable |
| GFI > 0.9 | 0.987 | Acceptable |
| Estimated parameters are statistically significant | t > 1.96 | Acceptable |

**2.4 Results Analysis**

Based on the empirical analysis described above, the following conclusions can be drawn by testing the hypotheses:

(1) In the context of the structural equation model referenced in Figure 3, the numbers presented (0.54, 0.81, 0.88, 0.79, 0.68, and 0.81) are standardized path coefficients. These coefficients measure the strength and direct influence of each factor—regulatory compliance, green procurement, green manufacturing, green logistics, green sales, and internal environmental protection and cost control—on the level of recognition of green supply chain management (GSCM) practices within chemical enterprises. A higher coefficient value indicates a greater positive impact of the factor on GSCM recognition. For instance, a coefficient of 0.81 suggests a strong positive influence of green procurement on the recognition of GSCM, implying that when enterprises actively engage in environmentally responsible purchasing practices, their awareness and valuation of GSCM principles are likely to be higher. These coefficients serve as critical indicators for assessing which aspects of green supply chain management are most effectively recognized within the industry and can guide future strategic enhancements in GSCM practices. This indicates that green procurement, green manufacturing, green logistics, green sales, and internal environmental protection and cost control have a significant positive impact on the recognition of GSCM. Well-established regulations also play a promotional role in the recognition of GSCM. The direct effects of the same factors on the implementation of GSCM are 0.57, 0.86, 0.87, 0.75, 0.77, and 0.81, respectively. This demonstrates that green procurement, green manufacturing, green logistics, green sales, and internal environmental protection and cost control significantly positively influence the implementation of GSCM, and that well-established regulations also enhance the implementation of GSCM.

(2) According to Figure 3, the direct effects of the recognition and implementation of GSCM on economic and environmental benefits are 0.60 and 0.91, respectively. This suggests that the implementation of GSCM has a significant positive impact on the economic and environmental benefits of enterprises, and that the recognition of GSCM also greatly promotes the improvement of these benefits.

(3) Comparing the magnitude of the direct effects of each link, improvements in the five stages of green procurement, green manufacturing, green logistics, green sales, and internal environmental protection and cost control will greatly propel the practice of GSCM in enterprises. The perfection of regulations also plays a promotional role in the practice of GSCM. Therefore, in the chemical industry cluster, to optimize the green supply chain management system, the focus should be on enhancing green procurement, manufacturing, logistics, sales, and internal environmental protection and cost control. Additionally, the government should pay attention to improving the legal framework for environmental protection.

## 3 Optimization of Green Supply Chain Management System for the Chemical Industry Cluster

### 3.1 Establishing a Comprehensive Policy and Regulatory Framework

The analysis of the survey indicates that emission reduction policies in the chemical industry are often in the form of plans with weak enforcement and insufficient impact, failing to achieve the anticipated effects of the policies. Firstly, the government should propose actionable legislation, with detailed inspection indicators to ensure the laws are effectively implemented. Furthermore, the severe pollution issues of the chemical industry should receive greater attention from higher-level government departments to enhance the impact and enforcement of policies [39].

### 3.2 Process Optimization

According to the survey, the chemical industry generates a significant number of pollutants such as wastewater, emissions, and solid waste during production, which severely impacts the environment. To reduce costs and environmental pollution, enterprises should first strengthen the supervision mechanisms for suppliers within green procurement. Following this, establishing long-term stable relationships with suppliers for technological innovation can reduce purchasing costs as well as the costs of production, manufacturing, and waste disposal. To lower costs and increase safety during manufacturing and transportation, the chemical industry should focus more on upgrading production equipment and technological innovation. Meanwhile, the government should actively promote green logistics, establish relevant policies, regulations, and standards, and increase the construction of logistics infrastructure. Additionally, the concept of green sales is not widely understood by enterprises and consumers at present, so the chemical industry needs to continuously improve the efficiency of green sales implementation through establishing corporate green marketing concepts and transforming and upgrading technological processes.

### 3.3 Adjusting the Chemical Industry Structure

The use of high technology in the chemical industry has become a trend. The new model for the chemical industry aims to use advanced technology to address ecological and environmental damage caused by chemical processes, achieving harmonious and sustainable development of the economy and the environment. China needs to introduce subsidy policies to encourage technological innovation in the chemical industry, benefiting the environment and society.

### 3.4 Leveraging Scale Effects of Resource Utilization and Waste Disposal within the Cluster

The chemical industry can realize the scale effects of resource utilization and waste disposal within the cluster only by closely linking with upstream and downstream enterprises to share resources and technology, forming chemical industrial parks[40]. By rational coordination of raw materials and finished products and mutually utilizing the resources of various enterprises, the chemical industry can lower production costs and waste disposal expenses, thereby enhancing economic and environmental benefits.

## 4 Conclusion

This paper surveyed the current state of green supply chain management in the chemical industry, using exploratory factor analysis and constructing an AMOS structural equation model to study the main factors and effect rates influencing the practice of green supply chain management. Recommendations for improvement have been proposed in areas such as refining regulations, raw material procurement, product manufacturing, logistics and transportation, product sales, promotion of technological innovation, and resource recycling. These suggestions aim to provide references for the chemical industry to advance green innovation and sustainable development.

## 5.Reference

[1]     T.-L. Chen, H. Kim, S.-Y. Pan, P.-C. Tseng, Y.-P. Lin, and P.-C. Chiang, "Implementation of green chemistry principles in circular economy system towards sustainable development goals: Challenges and perspectives," *Science of the Total Environment,* vol. 716, p. 136998, 2020.

[2]     X. Deng, S. Oda, and Y. Kawano, "Split-joint bull's eye structure with aperture optimization for multi-frequency terahertz plasmonic antennas," in *2016 41st International Conference on Infrared, Millimeter, and Terahertz waves (IRMMW-THz)*, 2016: IEEE, pp. 1-2.

[3]     H. Desing, D. Brunner, F. Takacs, S. Nahrath, K. Frankenberger, and R. Hischier, "A circular economy within the planetary boundaries: Towards a resource-based, systemic approach," *Resources, Conservation and Recycling,* vol. 155, p. 104673, 2020.

[4]     F. Yu and J. Strobel, "Work-in-Progress: Pre-college Teachers' Metaphorical Beliefs about Engineering," in *2021 IEEE Global Engineering Education Conference (EDUCON)*, 2021: IEEE, pp. 1497-1501.

[5]     M. Negri, E. Cagno, C. Colicchia, and J. Sarkis, "Integrating sustainability and resilience in the supply chain: A systematic literature review and a research agenda," *Business Strategy and the environment,* vol. 30, no. 7, pp. 2858-2886, 2021.

[6]     G. Sun, T. Zhan, B. G. Owusu, A.-M. Daniel, G. Liu, and W. Jiang, "Revised reinforcement learning based on anchor graph hashing for autonomous cell activation in cloud-RANs," *Future Generation Computer Systems,* vol. 104, pp. 60-73, 2020.

[7]     X. Deng, L. Li, M. Enomoto, and Y. Kawano, "Continuously frequency-tuneable plasmonic structures for terahertz bio-sensing and spectroscopy," *Scientific reports,* vol. 9, no. 1, p. 3498, 2019.

[8]     Y. Shen, H.-m. Gu, L. Zhai, B. Wang, S. Qin, and D.-w. Zhang, "The role of hepatic Surf4 in

lipoprotein metabolism and the development of atherosclerosis in apoE−/− mice," *Biochimica et Biophysica Acta (BBA)-Molecular and Cell Biology of Lipids,* vol. 1867, no. 10, p. 159196, 2022.

[9] B. Wang *et al.*, "Atherosclerosis-associated hepatic secretion of VLDL but not PCSK9 is dependent on cargo receptor protein Surf4," *Journal of Lipid Research,* vol. 62, 2021.

[10] S.-j. Deng, Y. Shen, H.-M. Gu, S. Guo, S.-R. Wu, and D.-w. Zhang, "The role of the C-terminal domain of PCSK9 and SEC24 isoforms in PCSK9 secretion," *Biochimica et Biophysica Acta (BBA)-Molecular and Cell Biology of Lipids,* vol. 1865, no. 6, p. 158660, 2020.

[11] Y. Shen *et al.*, "Surf4 regulates expression of proprotein convertase subtilisin/kexin type 9 (PCSK9) but is not required for PCSK9 secretion in cultured human hepatocytes," *Biochimica et Biophysica Acta (BBA)-Molecular and Cell Biology of Lipids,* vol. 1865, no. 2, p. 158555, 2020.

[12] D. Xia, A. K. Alexander, A. Isbell, S. Zhang, J. Ou, and X. M. Liu, "Establishing a co-culture system for Clostridium cellulovorans and Clostridium aceticum for high efficiency biomass transformation," *J. Sci. Heal. Univ. Ala,* vol. 14, pp. 8-13, 2017.

[13] Y. Shen, H.-M. Gu, S. Qin, and D.-W. Zhang, "Surf4, cargo trafficking, lipid metabolism, and therapeutic implications," *Journal of Molecular Cell Biology,* vol. 14, no. 9, p. mjac063, 2022.

[14] M. Wang *et al.*, "Identification of amino acid residues in the MT-loop of MT1-MMP critical for its ability to cleave low-density lipoprotein receptor," *Frontiers in Cardiovascular Medicine,* vol. 9, p. 917238, 2022.

[15] W. Dai, "Evaluation and Improvement of Carrying Capacity of a Traffic System," *Innovations in Applied Engineering and Technology,* pp. 1-9, 2022.

[16] W. Dai, "Safety evaluation of traffic system with historical data based on Markov process and deep-reinforcement learning," *Journal of Computational Methods in Engineering Applications,* pp. 1-14, 2021.

[17] V. S. Narwane, R. D. Raut, V. S. Yadav, N. Cheikhrouhou, B. E. Narkhede, and P. Priyadarshinee, "The role of big data for Supply Chain 4.0 in manufacturing organisations of developing countries," *Journal of enterprise information management,* vol. 34, no. 5, pp. 1452-1480, 2021.

[18] S. Kundu, Y. Fu, B. Ye, P. A. Beerel, and M. Pedram, "Toward Adversary-aware Non-iterative Model Pruning through D ynamic N etwork R ewiring of DNNs," *ACM Transactions on Embedded Computing Systems,* vol. 21, no. 5, pp. 1-24, 2022.

[19] X. Deng, S. Oda, and Y. Kawano, "Frequency selective, high transmission spiral terahertz plasmonic antennas," *Journal of Modeling and Simulation of Antennas and Propagation,* vol. 2, pp. 1-6, 2016.

[20] Y. Liu, M. Hajj, and Y. Bao, "Review of robot-based damage assessment for offshore wind turbines," *Renewable and Sustainable Energy Reviews,* vol. 158, p. 112187, 2022.

[21] Y. Qiu, "Estimation of tail risk measures in finance: Approaches to extreme value mixture modeling," Johns Hopkins University, 2019.

[22] Y. Liu and Y. Bao, "Review on automated condition assessment of pipelines with machine learning," *Advanced Engineering Informatics,* vol. 53, p. 101687, 2022.

[23] X. Deng, Z. Dong, X. Ma, H. Wu, and B. Wang, "Active gear-based approach mechanism for scanning tunneling microscope," in *2009 International Conference on Mechatronics and Automation*, 2009: IEEE, pp. 1317-1321.

[24] T. Sugaya and X. Deng, "Resonant frequency tuning of terahertz plasmonic structures based on solid immersion method," in *2019 44th International Conference on Infrared, Millimeter, and Terahertz Waves (IRMMW-THz)*, 2019: IEEE, pp. 1-2.

[25] Z. Luo, H. Xu, and F. Chen, "Audio Sentiment Analysis by Heterogeneous Signal Features Learned from Utterance-Based Parallel Neural Network," in *AffCon@ AAAI*, 2019: Shanghai, China, pp. 80-87.

[26] F. Chen, Z. Luo, Y. Xu, and D. Ke, "Complementary fusion of multi-features and multi-modalities in sentiment analysis," *arXiv preprint arXiv:1904.08138,* 2019.

[27] Z. Luo, X. Zeng, Z. Bao, and M. Xu, "Deep learning-based strategy for macromolecules classification with imbalanced data from cellular electron cryotomography," in *2019 International Joint Conference on Neural Networks (IJCNN)*, 2019: IEEE, pp. 1-8.

[28] J. Horne *et al.*, "Caffeine and Theophylline Inhibit β-Galactosidase Activity and Reduce Expression in Escherichia coli," *ACS omega,* vol. 5, no. 50, pp. 32250-32255, 2020.

[29] M. B. Mock, S. Zhang, B. Pniak, N. Belt, M. Witherspoon, and R. M. Summers, "Substrate promiscuity of the NdmCDE N7-demethylase enzyme complex," *Biotechnology Notes,* vol. 2, pp. 18-25, 2021.

[30] F. Yu, J. Milord, S. L. Orton, L. Flores, and R. Marra, "The concerns and perceived challenges students faced when traditional in-person engineering courses suddenly transitioned to remote learning," in *2022 ASEE Annual Conference*, 2022.

[31] J. Milord, F. Yu, S. Orton, L. Flores, and R. Marra, "Impact of COVID Transition to Remote Learning on Engineering Self-Efficacy and Outcome Expectations," in *2021 ASEE Virtual Annual Conference*, 2021.

[32] F. Yu, J. O. Milord, L. Y. Flores, and R. Marra, "Work in Progress: Faculty choice and reflection on teaching strategies to improve engineering self-efficacy," in *2022 ASEE Annual Conference*, 2022.

[33] F. Yu, J. Milord, S. Orton, L. Flores, and R. Marra, "Students' Evaluation Toward Online Teaching Strategies for Engineering Courses during COVID," in *2021 ASEE Midwest Section Conference*, 2021.

[34] S. Li, K. Singh, N. Riedel, F. Yu, and I. Jahnke, "Digital learning experience design and research of a self-paced online course for risk-based inspection of food imports," *Food Control,* vol. 135, p. 108698, 2022.

[35] F. Chen and Z. Luo, "Learning robust heterogeneous signal features from parallel neural network for audio sentiment analysis," *arXiv preprint arXiv:1811.08065,* 2018.

[36] Z. Luo, H. Xu, and F. Chen, "Utterance-based audio sentiment analysis learned by a parallel combination of cnn and lstm," *arXiv preprint arXiv:1811.08065,* 2018.

[37] L. Pinto, "Green supply chain practices and company performance in Portuguese manufacturing sector," *Business Strategy and the Environment,* vol. 29, no. 5, pp. 1832-1849, 2020.

[38] S. Gold and M. C. Schleper, "A pathway towards true sustainability: A recognition foundation of sustainable supply chain management," *European Management Journal,* vol. 35, no. 4, pp. 425-429, 2017.

[39]  J. Wu, M. Xu, and P. Zhang, "The impacts of governmental performance assessment policy and citizen participation on improving environmental performance across Chinese provinces," *Journal of Cleaner Production,* vol. 184, pp. 227-238, 2018.

[40]  K. Liu, X. Wang, and Y. Yan, "Network analysis of industrial symbiosis in chemical industrial parks: A case study of Nanjing Jiangbei new materials high-tech park," *Sustainability,* vol. 14, no. 3, p. 1381, 2022.